\documentclass[twocolumn,prb,reprint,superscriptaddress,floatfix,amsmath,amssymb,aps]{revtex4-2}
\pdfoutput=1

\usepackage[utf8]{inputenc}
\usepackage[caption=false,subrefformat=parens,labelformat=parens]{subfig}
\usepackage{graphicx}
\usepackage{float}
\usepackage{amsmath}
\usepackage{dcolumn}
\usepackage{bm}
\usepackage{xcolor}
\usepackage{xfrac}
\usepackage{chemformula}
\usepackage{siunitx}
\usepackage{booktabs}
\usepackage{braket}

\usepackage{tabularx}
\usepackage{amsfonts}
\usepackage{amsmath}
\usepackage{amssymb}
\usepackage{bm}
\usepackage{graphicx}
\usepackage{dcolumn}
\usepackage{mciteplus}
\usepackage{euscript}
\usepackage{multirow}
\usepackage{color}
\usepackage{textcomp}
\usepackage{gensymb}
\usepackage{float}
\usepackage{enumitem}
\usepackage{xcolor}
\usepackage{sepfootnotes}
\usepackage{mhchem}
\usepackage{outlines}

\newcommand{\ef}{E_{\rm F}}

\newcommand{\rfig}[1]{Fig.~\ref{#1}}
\newcommand{\rFig}[1]{Figure~\ref{#1}}

\begin{document}
\title{Quantum Oscillations Measurement of the Heavy Electron Mass near the van Hove Singularity of a Kagome Metal}
\author{Elliott Rosenberg}
\affiliation{Department of Physics, University of Washington, Seattle, WA 98112, USA}
\author{Jonathan M. DeStefano}
\affiliation{Department of Physics, University of Washington, Seattle, WA 98112, USA}
\author{Yongbin Lee}
\affiliation{Ames Laboratory, U.S. Department of Energy, Ames, Iowa 50011, USA}
\author{Chaowei Hu}
\affiliation{Department of Physics, University of Washington, Seattle, WA 98112, USA}
\author{Yue Shi}
\affiliation{Department of Physics, University of Washington, Seattle, WA 98112, USA}
\author{David Graf}
\affiliation{National High Magnetic Field Laboratory, Tallahassee, FL 32310, USA}
\author{Shermane M. Benjamin}
\affiliation{National High Magnetic Field Laboratory, Tallahassee, FL 32310, USA}
\author{Liqin Ke}
\affiliation{Ames Laboratory, U.S. Department of Energy, Ames, Iowa 50011, USA}
\author{Jiun-Haw Chu}
\affiliation{Department of Physics, University of Washington, Seattle, WA 98112, USA}
\date{\today}

\begin{abstract}
Kagome metals with the Fermi energy tuned near the van Hove singularities (vHss) have shown to host exotic phases including unconventional superconductivity and a chiral flux phase arising from a charge density wave. However, most quantum oscillations studies of the electronic structure of kagome metals focus on compounds which electronically or magnetically order, obscuring the unperturbed vHs. Here we present quantum oscillation measurements of \ch{YV6Sn6} which contains a pristine kagome lattice free from long range order. We discovered quantum oscillations corresponding to a large orbit ($\approx$70\% of the Brillouin Zone area) with the heaviest mass ever observed in vanadium based kagome metals ($\approx3.3 m_e$), consistent with a Fermi pocket whose Fermi level is near the vHs. Comparing with first principle calculations suggests that the effective mass of this pocket is highly sensitive to the position of Fermi level. Our study establishes the enhanced density of states associated with a vHs in a kagome metal, allowing further insight into a potential driving mechanism for the unconventional electronic orderings in this class of materials.
\end{abstract}

\maketitle

\section{Introduction}
Materials with Fermi energies tuned near or at flat bands or van Hove singularities (vHss) are extensively studied, as they can host a variety of electronic instabilities due to the enhanced density of states. Recently, kagome lattices have received significant attention as tight binding model calculations reveal vHss near the Fermi level, as well as flat bands and Dirac points in the electronic band structure of the host materials regardless of which species of atoms are in the lattice~\cite{guoTopologicalInsulatorKagome2009, mazinTheoreticalPredictionStrongly2014, chenAnomalousHallEffect2014, tangHighTemperatureFractionalQuantum2011, kangDiracFermionsFlat2020}. While kagome materials with flat bands at the Fermi energy remain rare/unexplored, several families have been recently investigated for which the Fermi energy is near vHss, most notably the \textit{A}\ch{V3Sb5} (\textit{A} = K, Rb, Cs) superconductor family, and it is believed the enhanced density of states plays a central role in the hosting the charge density wave (CDW) phases and unconventional superconductivity for these materials~\cite{wu2021nature,lin2021complex}.

Van Hove singularities, which are saddle points in the electronic band structure that often occur at the Brillouin zone (BZ) boundary and are accompanied by a divergence of the effective mass and density of states, appear in a variety of notable materials. They play a prominent role in high-$T_c$ superconductors like the cuprates~\cite{markiewicz1993van,piriou2011first,tsuei1990anomalous}, and are believed to host potentially exotic superconductivity in \ch{Sr2RuO4}~\cite{lu1996fermi,wang2013theory}. Recently, graphene, which has a honeycomb lattice that produces a very similar band structure as the kagome lattice, has been tuned through van Hove singularities in twisted bilayers via electrostatic gating~\cite{li2010observation}. Computations predict the stabilization of a unified charge instability and nearby superconductivity~\cite{mcchesney2010extended,sherkunov2018electronic}. Thus materials containing kagome lattices, which are theoretically guaranteed to host vHss if only nearest neighbor hopping terms are included in a tight-binding model, provide a promising playground to investigate the properties and exotic phases driven by these band structure features. 


In particular the \textit{A}\ch{V3Sb5} family, which has a pristine kagome lattice comprised of only V ions, has gained significant attention due to the combination of a high temperature ($\approx$\SI{80}{K}) charge density wave (CDW) and a low temperature ($<$\SI{3}{K}) superconducting state~\cite{KVS_SC, RVS_props, CVS_correlated}. The CDW is of particular interest due to a variety of observed orderings including $2\times2\times2$~\cite{AVS_222} and $2\times2\times4$~\cite{AVS_224}, and claims that it may break time-reversal symmetry~\cite{TRSbreaking_kagome, saykin2023high}. However, even after extensive investigation into the nature of the CDW phase, its origin remains unclear, as does the extent the vHs from the kagome band structure contributes in driving it. A commensurate $2\times2$ lattice distortion implies Fermi surface nesting may play a key role in driving the CDW, but Raman measurements have pointed to strong electron-phonon coupling likely being a key component in \ch{CsV3Sb5}~\cite{CVS_epcoupling, CVS_ampmodes}. 

Another kagome material with recently discovered exotic CDW phases is the antiferromagnet FeGe, where magnetic Fe ions comprise the kagome lattice. Below \SI{100}{K}, neutron scattering measurements observed peaks with the same wave vector as seen in the \textit{A}\ch{V3Sb5} family~\cite{FeGe_CDW}. A concurrent anomalous Hall effect (AHE) at magnetic fields above the spin-flop transition was also observed, potentially arising from chiral loop currents. It is proposed that the van Hove singularities of the kagome bands provide the increased density of states to host this CDW phase~\cite{FeGe_CDW}. This finding has raised questions of the nature of kagome band structures which host exotic CDW states, as well as the role of the magnetism of the kagome ions themselves in the appearance of superconductivity, as only the non-magnetic V lattice exhibits superconductivity.

The \textit{R}\ch{V6Sn6} family contains a pristine kagome lattice composed only of vanadium ions, analogous to the \textit{A}\ch{V3Sb5} family, but also allows a magnetic (non-magnetic) tuning parameter through the substitution of the \textit{R} ion, from Sm-Yb (Lu, Y, and Sc). Notably, \ch{ScV6Sn6} was recently synthesized and found to exhibit a CDW at roughly \SI{90}{K}~\cite{SVSdiscovery}. There is evidence that this CDW also exhibits signatures of time-reversal symmetry breaking through an apparent AHE~\cite{mozaffari2023universal,yi2023charge}. However, early studies have found distinct differences between \ch{ScV6Sn6} and the \textit{A}\ch{V3Sb5} family - no superconductivity has been found in \ch{ScV6Sn6} down to \SI{40}{mK} even under high pressures~\cite{SVSpressure}, in \ch{ScV6Sn6} the reconstruction accompanying the CDW is the in-plane $3\times3$ as opposed to $2\times2$~\cite{SVSdiscovery}, and in \ch{ScV6Sn6} the lattice distortion associated with the CDW is mostly along the c-axis~\cite{SVSdiscovery} whereas the distortion in \textit{A}\ch{V3Sb5} is mostly in the ab-plane~\cite{AVS_224}. 

 
It is thus an open question as to the universality of the vanadium kagome lattice in hosting exotic CDWs, and the extent to which the putative van Hove singularities supply the density of states to host/drive these phases. Measurements such as quantum oscillations which reveal the size and effective masses of carrier pockets will be crucial in determining the existence/characteristics of a vHs in these materials. However, to measure the unperturbed/non-reconstructed vanadium band structure, only YV$_6$Sn$_6$ and LuV$_6$Sn$_6$ are candidate materials as the rest of the members of the \textit{A}\ch{V3Sb5} and \textit{R}\ch{V6Sn6} families either undergo reconstruction from a CDW phase or have magnetic phase transitions which can potentially obscure results.  This motivates quantum oscillations measurements of \ch{YV6Sn6} at low temperatures.

\ch{YV6Sn6} has been reported recently as a member of the \textit{R}\ch{V6Sn6} family which does not undergo a phase transition down to \SI{1.8}{K}~\cite{pokharelElectronicPropertiesTopological2021}. The Y ion has an empty $4f$ shell and so contains no local magnetic or multipolar moments, rather acting as a spacer between the vanadium kagome lattices and hexagonal lattices of tin. Besides exhibiting nonlinear-two band Hall magnetotransport at temperatures below 20K, this material has a smooth resistivity and heat capacity with no evident magnetism observed in magnetization measurements~\cite{pokharelElectronicPropertiesTopological2021}. Thus this material allows low-temperature measurements to probe the undisturbed kagome band structure, which we predict via Density Functional Theory (DFT) calculations to have a vHs at or very near the Fermi energy, as shown in Figure \ref{fig:dft}a.  This paper displays the results of quantum oscillations of \ch{YV6Sn6} seen in torque magnetometry from an applied magnetic field up to 41.5 T, allowing multiple sets of frequencies to be identified including a relatively heavy high frequency which may belong to the vanadium kagome band containing a vHs near the Fermi energy.

\section{Results}
\subsection{Experimental Results}
\begin{figure*}
    \centering
    \includegraphics[width=1\textwidth]{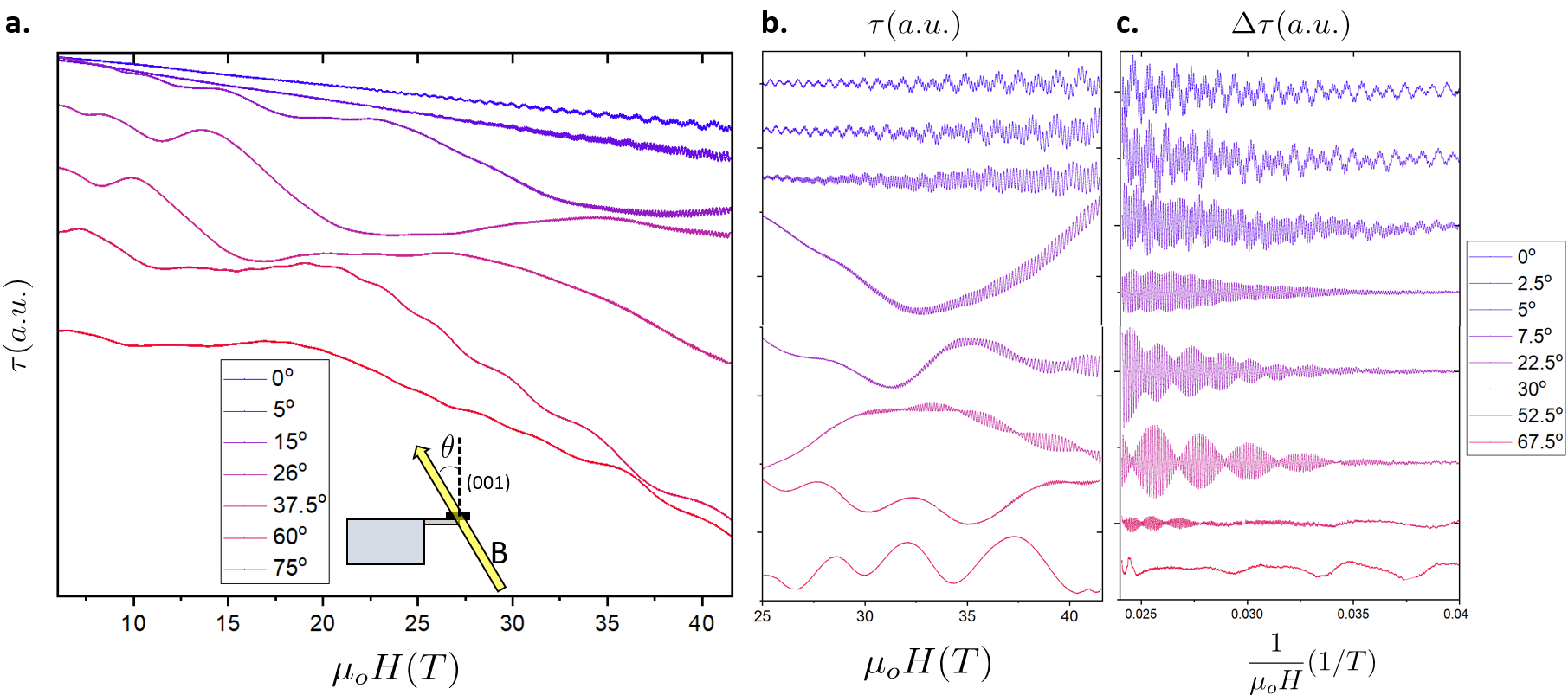}
    \caption{a) Raw torque magnetometry data (in arbitrary units) plotted for magnetic field at select angles (detailed in inset legend) offset for clarity. Inset schematic displays the experimental torque setup, with $\theta$ measuring the angle between the magnetic field and the sample's c-axis. b) Torque data with a linear background subtracted plotted against magnetic fields from 25 to 41.5 T, with offsets for clarity. The angles selected are shown in the legend on the right. c) Corresponding residual oscillations after a smooth spline background was subtracted, plotted against 1/$\mu_0H$.}
    \label{fig:rawtorque}
\end{figure*}

Figure \ref{fig:rawtorque} displays raw torque data vs. magnetic field from 6 T to 41.5 T, with selected traces representing data collected at different angles between the magnetic field and the c-axis of the sample. Note there is the co-existence of clear fast oscillations ($\approx$9000 T at $0^\circ$), medium-fast oscillations ($\approx$1000 T) which disappear at angles above $5^\circ$, and large amplitude slow oscillations ($\approx$40 T) which appear at angles above $15^\circ$ and below 25 T. This establishes this material as relatively unique in exhibiting the prominent coexistence of very slow and large oscillations with relatively high frequency oscillations, but also makes frequency extraction difficult as polynomial backgrounds will either obscure the slow or fast oscillations depending on the number of terms used. This was solved as detailed in the Materials and Methods section with an iterative spline algorithm with FFTs to isolate the high frequencies, and the low frequencies were determined manually by observing the spacing between the minimums and maximums of the signal with an optimized linear background.

\begin{figure*}
    \centering
    \includegraphics[width=1\textwidth]{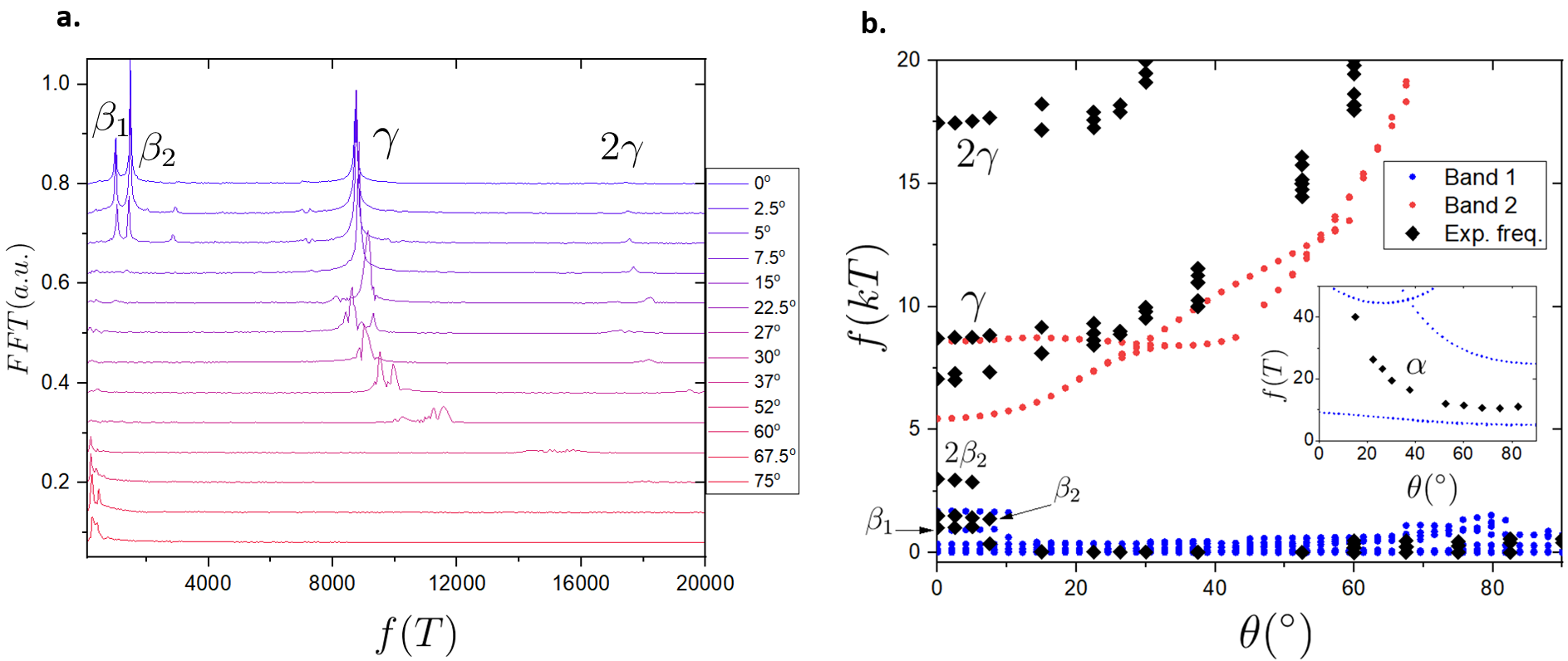}
    \caption{a) FFTs of the torque magnetometry data at different angles, offset for clarity. Only frequencies above 100 T are shown, as the low frequency was determined manually via the spacing of the apparent minimums and maximums of the data. The peaks are labeled by the pockets and harmonics they arise from, and their angular dependence of the peak maximums are shown as black diamonds in b) plotted against frequencies obtained from DFT calculations in color (matching the color scheme in Figure~\ref{fig:dft}.) The inset shows an expanded picture of the low frequency regime.}
    \label{fig:fft}
\end{figure*}
\begin{figure*}
    \centering
    \includegraphics[width=1\textwidth]{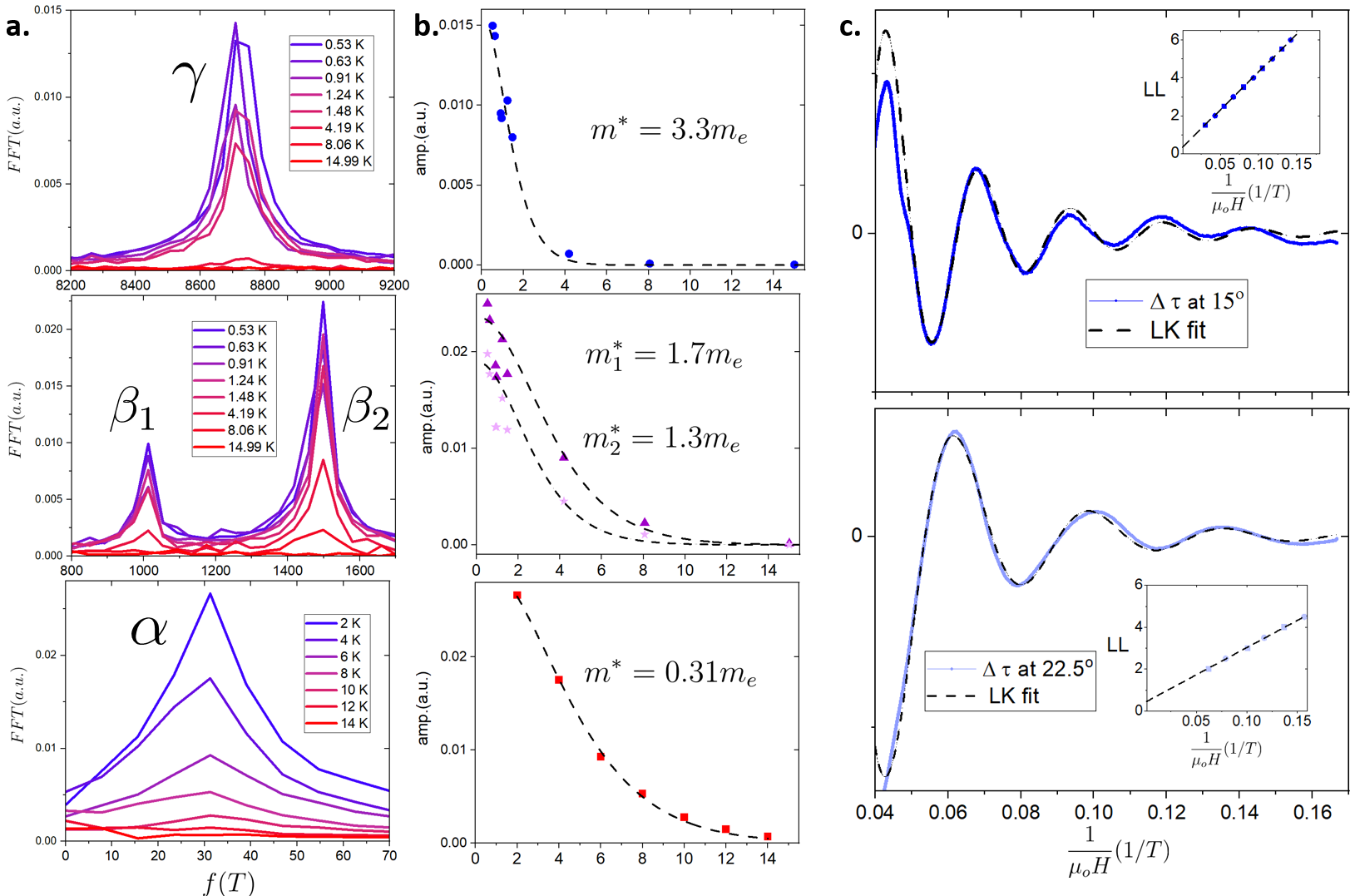}
    \caption{a) FFTs for different temperatures, with the orbit labeled above for each panel. b) Corresponding plot of the amplitudes against temperature, with Lifshitz-Kosevich temperature fits displayed with the black dashed lines. The extracted effective masses for each orbit are labeled in each of these panels. The FFT amplitudes for the top and middle panels were obtained with the magnetic field along the c-axis, whereas for the bottom panels it was obtained at 20 degrees from the c-axis. c) Plots of torque with a linear background subtracted against inverse field, for angles 15 degrees (top) and 22.5 degrees (bottom). The dashed lines are Lifschitz-Kosevich field fits, described in the main text. The insets depict Landau level plots, with the assigned Landau levels denoted by torque maxima and minima, plotted against inverse field, with linear fits described in the main text.  }
    \label{fig:effmass}
\end{figure*}


Figure~\ref{fig:fft}a displays the results of the frequencies obtained in the fast Fourier transforms (FFTs) from 100 to 20000 T, with the total FFT from each angle being offset for clarity. In Figure~\ref{fig:fft}b the peaks of the FFTs vs. angle (along with calculations which will be discussed later) are displayed, with the frequencies believed to be arising from orbits around separate pockets labeled accordingly. Pockets $\beta_1$ and $\beta_2$ denote orbits with frequencies at 1000 T and 1400 T respectively which move towards each other as angle is increased before they quickly disappear above 5 degrees. As shown in Figure~\ref{fig:effmass} these pockets have effective masses of 1.7 and 1.3$m_e$ respectively. Pocket $\gamma$ has the highest frequency orbits, and the frequency evolves from 8740 T at 0 degrees, to splitting into multiple closely spaced peaks ranging up to 20000 T at 60 degrees, before being unable to observe or resolve at angles above. The angle dependence of these peaks fits well to $1/cos(\theta)$, indicating that $\gamma$ is a 2D-like pocket, although it clearly has dispersion in $k_z$ to explain the peak splitting. Interestingly, the second and even third harmonic of this frequency can be observed (see section B of Supplementary Materials), indicating the high quality of the sample and the long mean free path for these orbits. The effective mass of this pocket was found to be 3.3$m_e$ as shown in Figure~\ref{fig:effmass}a, the heaviest of any of the masses extracted from this material.

The inset of Figure~\ref{fig:fft}b depicts the frequencies of the slow large amplitude oscillations extracted manually, also corroborated with Lifshitz-Kosevich (LK) fits and Landau level plots shown in Figure~\ref{fig:effmass}c. These frequencies were extracted only from 6 to 25 T, as the amplitude of the oscillations is no longer observable above this field range. This is possibly due to reaching the quantum limit of the pocket as its frequency is extremely low, or could be attributed to a very slow beating effect from similar frequencies. A higher magnetic field range would be necessary to resolve between these two scenarios. Regardless, the area of the orbits around the $\alpha$ pocket(s) has an angle dependence which was well described by a $1/\sin(\theta)$ fit. This indicates an 3D tube-like shape of the pocket in which the cylinder axis is along the ab plane. The effective mass of this pocket was obtained to be  $\approx0.3m_e$. Other peaks were obtained in the FFTs near 70 degrees as shown at the bottom right of Figure~\ref{fig:fft}b, but cannot be definitively tied to any pockets.

Figure~\ref{fig:effmass}c displays LK fits for torque data at 15 degrees and 22.5 degrees plotted against inverse field, where the large amplitude slow oscillations were quite prominent from 6 to 25 T. Linear backgrounds were determined to minimize the error of these fits, as well as minimize the error of the linear fit of maxima/minima positions with Landau level numbers as shown in the insets of the top and bottom panels. These fits used the standard LK formula:

\begin{equation}
\tau\propto\frac{k_BT(\mu_oH)^{1/2}}{\sinh(\frac{14.69Tm^*}{\mu_{o}H})}\sin(2\pi(F/\mu_o H)+\phi)e^{-\pi m^*/e\hbar (\mu_o H)\tau}
\end{equation}

Where the first fraction term indicates the temperature dependence of the oscillation amplitude and the exponential term is the Dingle factor. The effective mass was taken as $0.31m_e$ from the temperature dependence of the FFT peak (measured at 20 degrees), and frequencies of 40 and 27 T were obtained respectively for 15 degrees and 22.5 degrees. The phases were determined to be $1.04\pi$ and $1.08\pi$, indicating potential non-trivial topology of the band. The Dingle temperature was found to be 3.8K, giving a scattering time of $\tau^{\alpha}_{sc} = 3.04\times 10^{-13}$s implying a mean free path $l_{\alpha}=400$ $\AA$. A similar analysis of the high frequency orbit was performed, giving a scattering time of $\tau^{\gamma}_{sc} = 4.75\times 10^{-13}$ s.
and free path $l_{\gamma}=860$ $\AA$.

\subsection{Computational Results}
\begin{figure*}
\centering
    \includegraphics[width=1\textwidth]{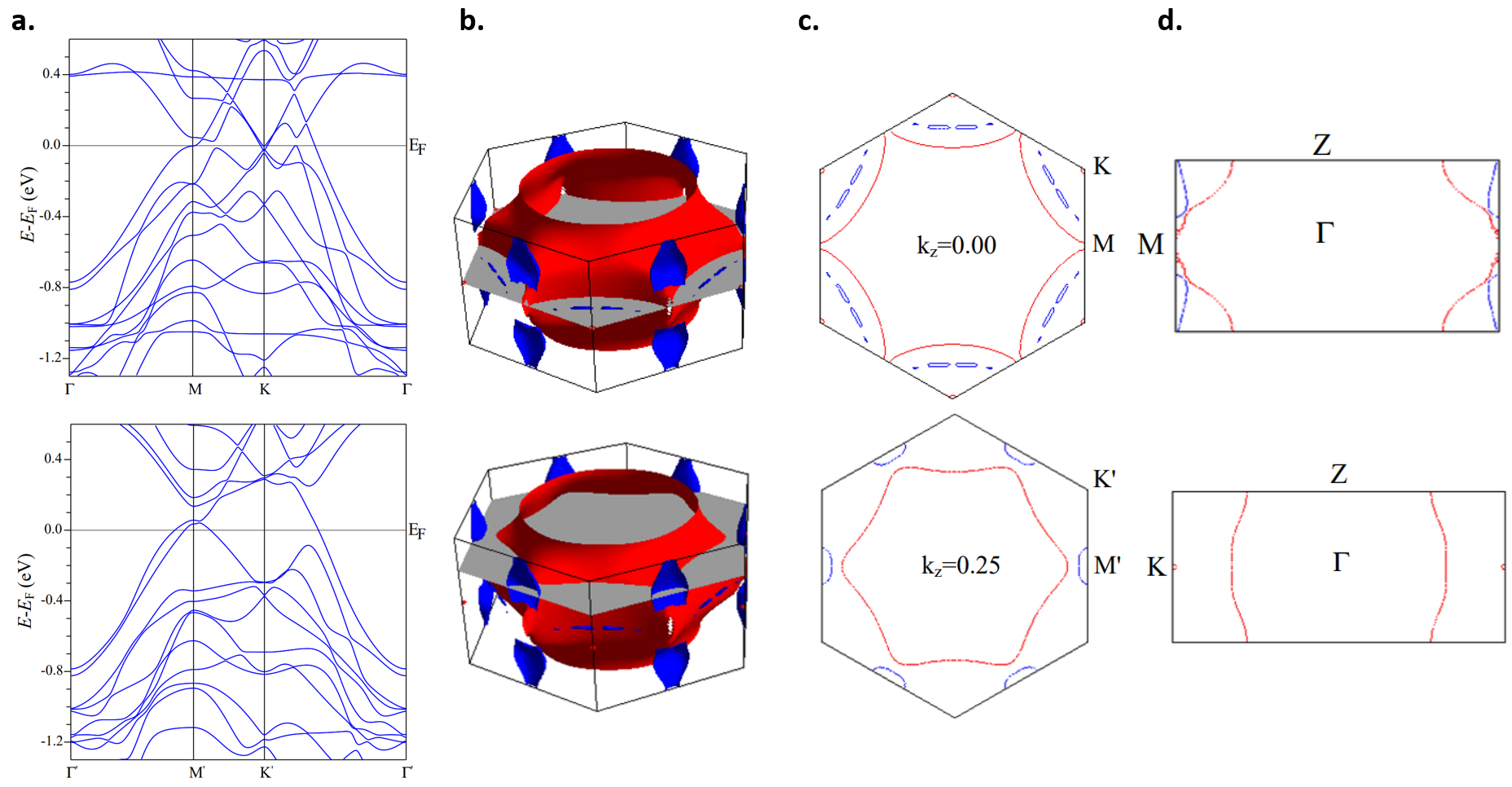}
    \caption{a) Band structures of \ch{YV6Sn6} calculated along a high-symmetry path at the planes of $k_{z}=0$ and $k_{z}=0.25$ (top and bottom). b) The 3D Fermi surface plot, along with two cross-section planes: (top row) corresponding to the (0 0 1) plane with $k_z = 0.0$, (bottom row) corresponding to the (0 0 1) plane with $k_z = 0.25$. c) High-symmetry $k$-points are marked in the cross-section plots for clarity. d) The Fermi surface with perpendicular cross-sections. Spin-orbit coupling is included in the calculations.}
    \label{fig:dft}
\end{figure*}

 Panels of \rFig{fig:dft}a show the band structure calculated with spin-orbit coupling along the high-symmetry $k$-path at two $k_z$ planes: (top) $k_{z} = 0$ and (bottom) $k_{z} = 0.25$. The band structures near the Fermi energy $\ef$ are dominated by V kagome lattices (see Supplementary Materials Figure S1 for sublattice-resolved partial DOS), exhibiting characteristic features of a kagome lattice, such as Dirac crossings (DCs) at the $K$-point and flat bands (FBs). There is also a vHs at the M point very close to the Fermi energy, as expected for a kagome lattice.
Unlike DCs, the flatness of FBs are \emph{not} topologically protected, and ideal FBs are only realized in the nearest-coupling limit. At $k_z=0$, as shown in the top panel of \rfig{fig:dft}a, two DCs occur right below $\ef$ at the $K$-point, and two flat-band-like features are found in this energy window, at approximately $0.4$ eV above and $1$ eV below E$_{\rm F}$, respectively.
However, the orbital characters of these DCs and FBs are \emph{not} dominated by more in-plane-like $d_{xy}$ and $d_{x^2-y^2}$ orbitals (see Supplementary Materials Figure S2 for V ${3d}$-resolved partial DOS). As a result, they show strong $k_z$ dependencies, as suggested by the difference between \rfig{fig:dft}a top and bottom. These findings are consistent with previous research on $R$Mn$_6$Sn$_6$ compounds, where most of the topological features at the Brillouin zone corner $K$ are strongly 3D, and the most pronounced quasi-2D dispersion is already too far above the $\ef$~\cite{lee2023prb}.

On the $k_{z} = 0$ plane, there is a large electron pocket centered at the $\Gamma$-point and a very small electron pocket centered at the $K$-point.
Additionally, several band crossings are located very close to $\ef$ on different $k$ points, generating small pieces of FS and contributing to the additional complexity of the FS.
On the $k_{z} = 0.25$ plane, two bands intersect the $\ef$, resulting in a large electron pocket centered at the $\Gamma$-point and a small hole pocket centered at the $M$-point.




Different colors in the Fermi surface plot represent different bands. The red FS sections correspond to electron pockets centered at $\Gamma$ and $K$, while the blue FS sections represent hole pockets mostly near $M'$. However, as shown in the left panel of Supplementary Materials Figure S3, the frequencies associated with the pockets do not quite match the observed experimental frequencies, most notably providing a 6 kT frequency orbit from the large red band as opposed to the experimentally observed 8.7kT. In an attempt to more accurately match the experimental frequencies, the Fermi surface was calculated with the Fermi energy shifted down by 3mRy (40meV). Although the results remain broadly similar, as demonstrated in comparing the left and right panels of Supplementary Materials Figure S3, the frequencies from the red electron band become larger while the small electron bands centered around the K points disappear. The orbits from the blue bands however display a more complicated frequency vs. angle relation, as more small pockets are present. The discussion will center on the comparison between the experimental frequencies and these shifted calculations. 

\subsection{Comparison between Experiments and Computations}
Our experimental findings are qualitatively consistent both with the DFT calculations presented in this work as well as with previous theoretical calculations of this compound and sister compounds \ch{TbV6Sn6} and \ch{GdV6Sn6}~\cite{pokharelElectronicPropertiesTopological2021,TVS_PRB}. The large frequency $\approx$ \SI{8700}{T} with $H||c$ observed can be matched quite closely with a calculated \SI{8570}{T} after the Fermi energy shift, also having a very similar angle dependence and splitting as shown in Figure~\ref{fig:fft}b in which the calculated and experimental frequencies are plotted together. The band that produces these orbits is calculated and shown in Figure~\ref{fig:dft}a, in which it has a minimum at $\Gamma$ and approaches a van Hove singularity at the M point very close to the Fermi level. The relatively heavy mass that was experimentally observed ($3.3m_e$) provides further evidence this orbit arises from where the band dispersion flattens out near the Fermi level. It is in fact heavier than the calculated effective mass ($2.2 m_e$ from Table I). It should be noted that the increase of calculated effective mass of the band with small energy shifts (comparing effective masses between the two sets of calculations) indicates the sensitivity of the density of states in the proximity of the van Hove singularity. This could potentially explain the discrepancy between the experimental and calculated effective mass, although we cannot rule out electronic correlations as ``dressing" the effective mass. The experimental splitting of this large pocket also indicates a relatively large $k_z$ dispersion, corroborated in the calculations displayed in Figure~\ref{fig:dft}a as the van Hove singularity is close to the Fermi energy for $k_z=0$ but farther away for $k_z=0.25$ (bottom panel). 
\begin{table}[htb]
  \label{tbl:eff-mass}  
  \caption{Experimental frequencies compared to the closest corresponding calculated frequencies for both shifted and unshifted DFT calculations.
  }
  \bgroup
  \def\arraystretch{1.3}
  \begin{tabularx}{\linewidth}{|*{3}{>{\centering\arraybackslash}X@{\vline}*{3}{>{\centering\arraybackslash}X|}}}
    \hline\hline
    \multicolumn{2}{|c|} {Experimental} & \multicolumn{2}{c|}{Calc. freq (0 shift)} & \multicolumn{2}{c|}{Calc. freq (shifted)} \\
    \hline \hline
    f(kT) & $m^*$  & f(kT) & $m^*$ & f(kT) & $m^*$  \\
    \hline
     8.7 & 3.3 & 5.89 & 1.2 & 8.59 & 2.2  \\
     \hline
     1.4 & 1.3  & 3.05 & 1.43 & 1.69 & 0.58  \\
    \hline
    1& 1.7  & --& -- & 0.91 & 0.93  \\
    \hline
     0.04 & 0.31 & 0.008 & 0.09 & 0.085 & 0.39  \\		
    \hline\hline
  \end{tabularx}
  \egroup
\end{table}

\begin{table}
  \centering
  \caption{Comparing \ch{YV6Sn6} and \ch{GdV6Sn6}}
  \label{tab:nice-table}
  \def\arraystretch{1.3}
  \begin{tabularx}{\linewidth}{|*{3}{>{\centering\arraybackslash}X@{\vline}*{3}{>{\centering\arraybackslash}X|}}}
    \hline
    \hline
    \multicolumn{3}{|c|}{YV$_6$Sn$_6$} & \multicolumn{3}{c|}{GdV$_6$Sn$_6$} \\
    \hline
    \hline
    f(kT) & {m$^*$/m$_e$} & $\tau(10^{-13}s)$ & {f(kT)} & m$^*$/m$_e$ &$ \tau(10^{-13}s)$ \\
    \hline
    0.040 & 0.31 & 3.04 & 0.092 & 0.58 & 1.35\\
    \hline
    1.01 & 1.7 & - & 0.85 & 1.15 & 1.08\\
    \hline
    1.50 & 1.4 & - & 1.46 & 0.88 & 0.57\\
    \hline
    8.72 & 3.3 & 4.75 & 8.28 & 2.25 & 2.82\\
    \hline
  \end{tabularx}
  \end{table}



The experimentally observed middle frequencies ($\beta_1= 1500T$ and $\beta_2 =1000T$) also have a correspondence with calculated frequencies, as shown in Figure~\ref{fig:fft}, in which calculated frequencies arising from the small blue bands qualitatively match both the angle dependence, quickly disappearing above 10 degrees, as well as the magnitudes of the frequencies. However as shown in Table I their measured effective masses are quite a bit heavier than the calculated ones. 

There are many expected low frequencies from the calculations, so it is difficult to accurately compare to the experimentally observed small pocket $\alpha$. As shown in the inset of Figure \ref{fig:fft}, there is no obvious calculated band that matches both the magnitude and angle dependence of the $\alpha$ pocket. For comparisons of observed and experimental frequencies in Table I, we take the band that produces frequencies with a similar 3D-like angle dependence $(\propto 1/sin(\theta))$ but with a larger frequency at 15 degrees ($\approx$\SI{80}{T}). We note in general it is consistent with the small and light electron pockets predicted to arise from the Dirac like dispersion near the K-points, especially given the potential non-trivial topology observed in the Berry phase of the orbit. 

The extracted band structure properties of \ch{YV6Sn6} can be compared to recently found results of its magnetic sister compound \ch{GdV6Sn6}~\cite{dhital2023evidence}. Quantum oscillations of \ch{GdV6Sn6} also reveal 4 major frequencies, shown in the right side of Table II, which are quite similar to the frequencies shown in this work for \ch{YV6Sn6}. We note the large pocket for \ch{YV6Sn6} appears to be slightly bigger as well as significantly heavier (3.3$m_e$ to 2.3$m_e$), providing evidence \ch{YV6Sn6} hosts a band with the vHs closer to the Fermi level. 

\section{Discussion}
This work, probing quantum oscillations measured by torque magnetometry, uncovered the heaviest/largest pocket seen so far in the vanadium based kagome family~\cite{dhital2023evidence,RVS_props,quantumtransport_CVS,SVS_AHE,zhang2022emergence,chen2022anomalous,yang2020giant,yi2024quantum,he2023quantum}. Corresponding DFT calculations of this material predict that the effective mass increases by almost twofold when the Fermi energy is shifted closer to the expected vHs to match the experimentally measured orbit area.   Hence, this comparison provides direct thermodynamic evidence of an enhanced density of states when the Fermi level is very close to a van Hove singularity of a kagome metal. 

Although we have established \ch{YV6Sn6} as an ideal system to study the van Hove singularity of an unperturbed pristine kagome lattice of vanadium ions, this compound remains unordered down to 0.3K, suggesting other parameters must be tuned to bring about a phase transition. Recent theoretical and experimental studies suggested that the Sb p-orbitals play an important role in the charge density wave formation in \textit{A}\ch{V3Sb5} family~\cite{Tsirlin2022,Jeong2022,Han2023,Ritz2023}. Similar analysis of the \ch{YV6Sn6} electronic structure may provide insights into the mechanism of electronic instabilities associated with the van Hove singularity in kagome metals.

\section{Materials and Methods}
\subsection{Crystal Growth}
Single crystals of \ch{YV6Sn6} were grown using a flux method similar to those previously reported~\cite{ishikawa2021gdv6sn6, leeAnisotropicMagneticProperty2022a, pokharelElectronicPropertiesTopological2021}. Y ingot (99.9\%), V powder (99.9\%), and Sn shot (99.999\%) were combined with atomic ratios 1:6:30, loaded into Canfield crucible sets~\cite{Canfield2016}, and vacuum-sealed in quartz tubes. These were heated to 1000$^\circ$C in 12 hours, held at this temperature for 24 hours, then cooled to 600$^\circ$C in 80 hours. The growths were then decanted in a centrifuge to separate the excess flux from the single crystals.
\subsection{Torque Magnetometry}
Small pieces of single crystals (50um x 50um x 20um) were attached using N-grease to resistive cantilevers of commercially available torque chips (SCL Sensor Tech PRSA-L300-F60-TL-CHP), and oriented such that the c-axis was perpendicular to the cantilever.  To isolate the signal arising from the bending of the cantilever a Wheatstone bridge was measured with resistors tuned to be close to the resistance of the cantilever at base temperature. The torque chips were attached to a rotator platform which could be externally controlled to rotate from 0$^\circ$ to 360$^\circ$, and loaded into the $^3$He system which can reach a base temperature of 0.3K and a maximum magnetic field of \SI{41.5}{T} at the National High Magnetic Field Laboratory in Tallahassee (Cell 6). 
\subsection{Quantum Oscillations Background Subtraction}
To construct backgrounds for the torque data that allowed accurate extraction of the higher frequencies ($>1000T$), an iterative spline algorithm was used, in which a smoothing spline with a very small second derivative penalty parameter (smoothing parameter) is first used as a background. It effectively acts a high pass filter, leaving only the high frequency oscillations invariant. A Fast Fourier Transform is performed, and the results saved. The background spline is then fed into this algorithm again, but now using a new smoothing spline with a larger smoothing parameter as the background, and the results of another FFT are added to the previous frequencies obtained. This process is iterated until all frequencies above a certain cutoff ($\approx 40T$, due to frequency binning constraints) were extracted. 

\subsection{Density Functional Theory Calculations}
Density Functional Theory calculations were performed using a full-potential linear augmented plane wave (FP-LAPW) method, as implemented in \textsc{wien2k}\cite{WIEN2k}.
Spin-orbit coupling (SOC) was included using a second variational method.
The generalized gradient approximation by Perdew, Burke, and Ernzerhof (PBE)~\cite{perdew1996} was used for the exchange-correlation potentials.
To generate the self-consistent potential and charge, we employed $R_\text{MT}\cdot K_\text{max}=8.0$ with Muffin-tin (MT) radii $R_\text{MT}=$ 2.5, 2.4, and 2.7 atomic units (a.u.)
for Y, V, and Sn, respectively.
Calculations found YV$_6$Sn$_6$ to be nonmagnetic, consistent with experiments that also found YV$_6$Sn$_6$ does not exhibit magnetic ordering down to \SI{0.3}{K}.
The primitive cell contains one formula unit, and experimental lattice parameters and atomic coordinates~\cite{romaka2011} are adopted in all calculations.
The self-consistent calculations were performed with 490 $k$-points in the irreducible Brillouin zone (IBZ) and were iterated until charge differences between consecutive iterations were smaller than $1\times10^{-3}e$, and the total energy difference was lower than 0.01 mRy.
After obtaining the self-consistent charge, band energies were calculated with a 64 $\times$ 64 $\times$ 33 fine k-mesh in the full Brillouin Zone (FBZ) for FS calculation.
We employed \textsc{FermiSurfer}~\cite{fermisurfer} to visualize the FS and \textsc{Skeaf}~\cite{skeaf} to calculate de Haas-van Alphen (dHvA) frequencies.

\section{Acknowledgements}

\textbf{Funding.} This work is supported by the Air Force Office of Scientific Research under grant FA9550-21-1-0068 and the David and Lucile Packard Foundation. This material is based upon work supported by the National Science Foundation Graduate Research Fellowship Program under Grant No. DGE-2140004. Any opinions, findings, and conclusions or recommendations expressed in this material are those of the authors and do not necessarily reflect the views of the National Science Foundation. L.K. and Y.L. are supported by the U.S.~Department of Energy, Office of Science, Office of Basic Energy Sciences, Materials Sciences and Engineering Division, and Early Career Research Program. Ames Laboratory is operated for the U.S.~Department of Energy by Iowa State University under Contract No.~DE-AC02-07CH11358. A portion of this work was performed at the National High Magnetic Field Laboratory, which is supported by the National Science Foundation (NSF) Cooperative Agreement No. DMR-1644779 and the State of Florida.

\textbf{Author Contributions.} J.-H.C. conceived of the experiment. E.R. and J.M.D. prepared the sample. E.R., J.M.D., C.H., Y.S., D.G., and S.M.B. performed the experiments. E.R. analyzed the data. Y.L. and L.K. performed the calculations. E.R., J.M.D. and J.-H.C. wrote the manuscript with input from all authors. 

\textbf{Competing Interests} The authors declare they have no competing interests.

\textbf{Data Availability} All data used to support the conclusions will be made available upon request.
\bibliography{main}
\newpage

\end{document}


\title{Supplementary Materials for ``Quantum Oscillations Measurement of the Heavy Electron Mass near the van Hove Singularity of a Kagome Metal"}
\author{Elliott Rosenberg}
\affiliation{Department of Physics, University of Washington, Seattle, WA 98112, USA}
\author{Jonathan M. DeStefano}
\affiliation{Department of Physics, University of Washington, Seattle, WA 98112, USA}
\author{Yongbin Lee}
\affiliation{Ames Laboratory, U.S. Department of Energy, Ames, Iowa 50011, USA}
\author{Chaowei Hu}
\affiliation{Department of Physics, University of Washington, Seattle, WA 98112, USA}
\author{Yue Shi}
\affiliation{Department of Physics, University of Washington, Seattle, WA 98112, USA}
\author{David Graf}
\affiliation{National High Magnetic Field Laboratory, Tallahassee, FL 32310, USA}
\author{Shermane M. Benjamin}
\affiliation{National High Magnetic Field Laboratory, Tallahassee, FL 32310, USA}
\author{Liqin Ke}
\affiliation{Ames Laboratory, U.S. Department of Energy, Ames, Iowa 50011, USA}
\author{Jiun-Haw Chu}
\affiliation{Department of Physics, University of Washington, Seattle, WA 98112, USA}

\date{\today}

\maketitle
\subsection{Calculations}

\begin{figure}[htb]
  \centering
  \begin{tabular}{c c}
    \includegraphics[width=.49\linewidth,clip]{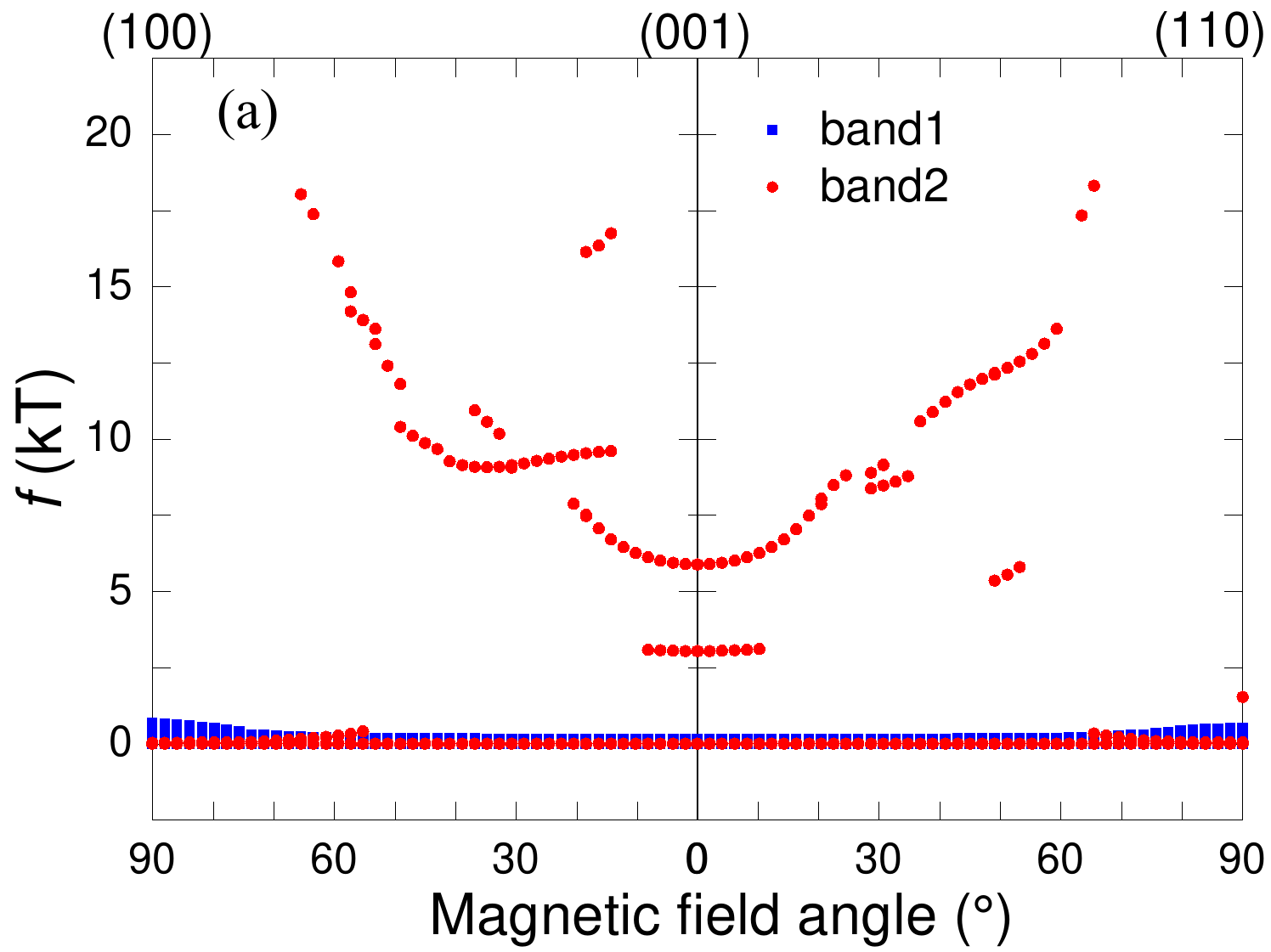} & \includegraphics[width=.49\linewidth,clip]{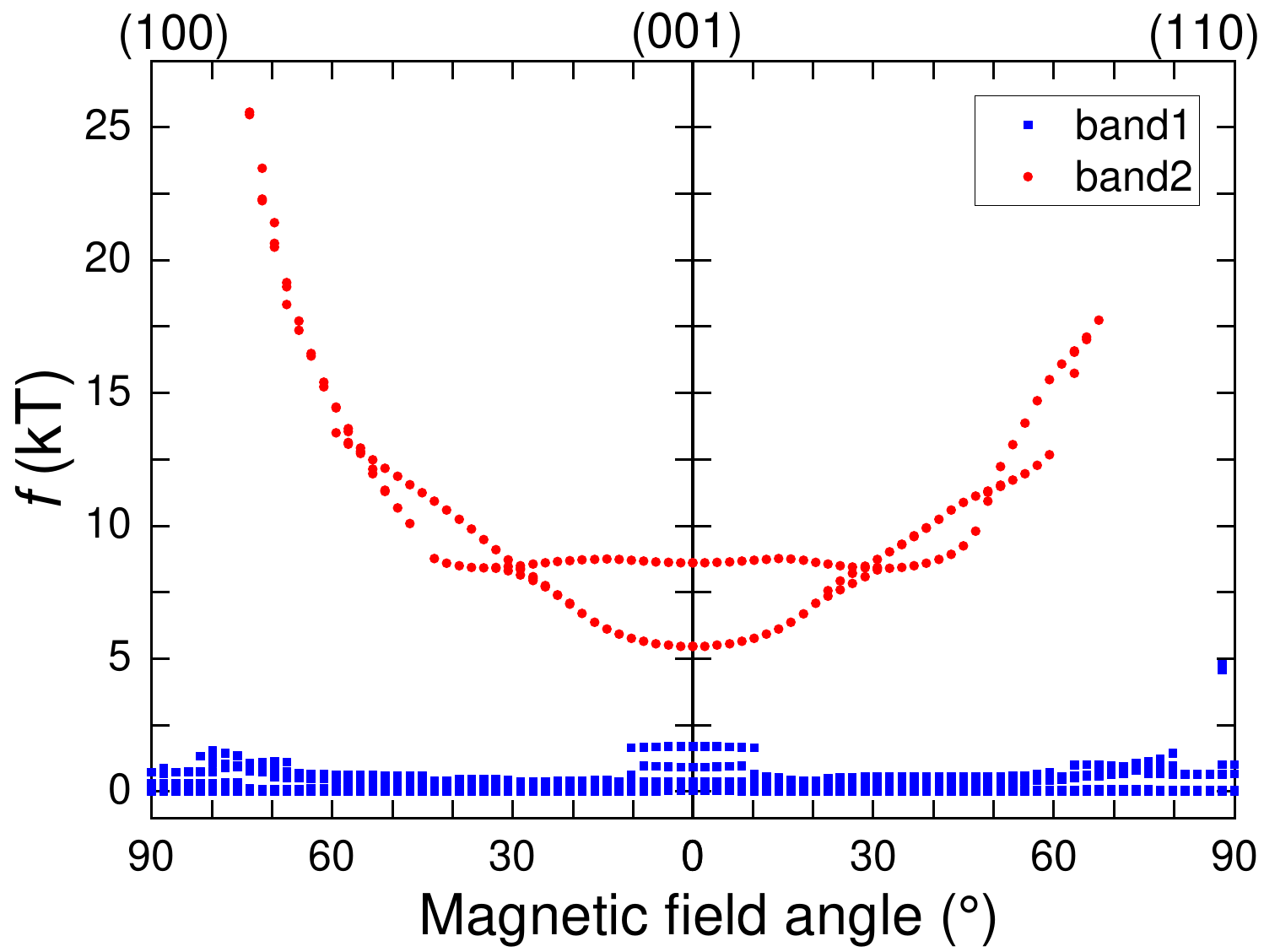} \\
    \includegraphics[width=.49\linewidth,clip]{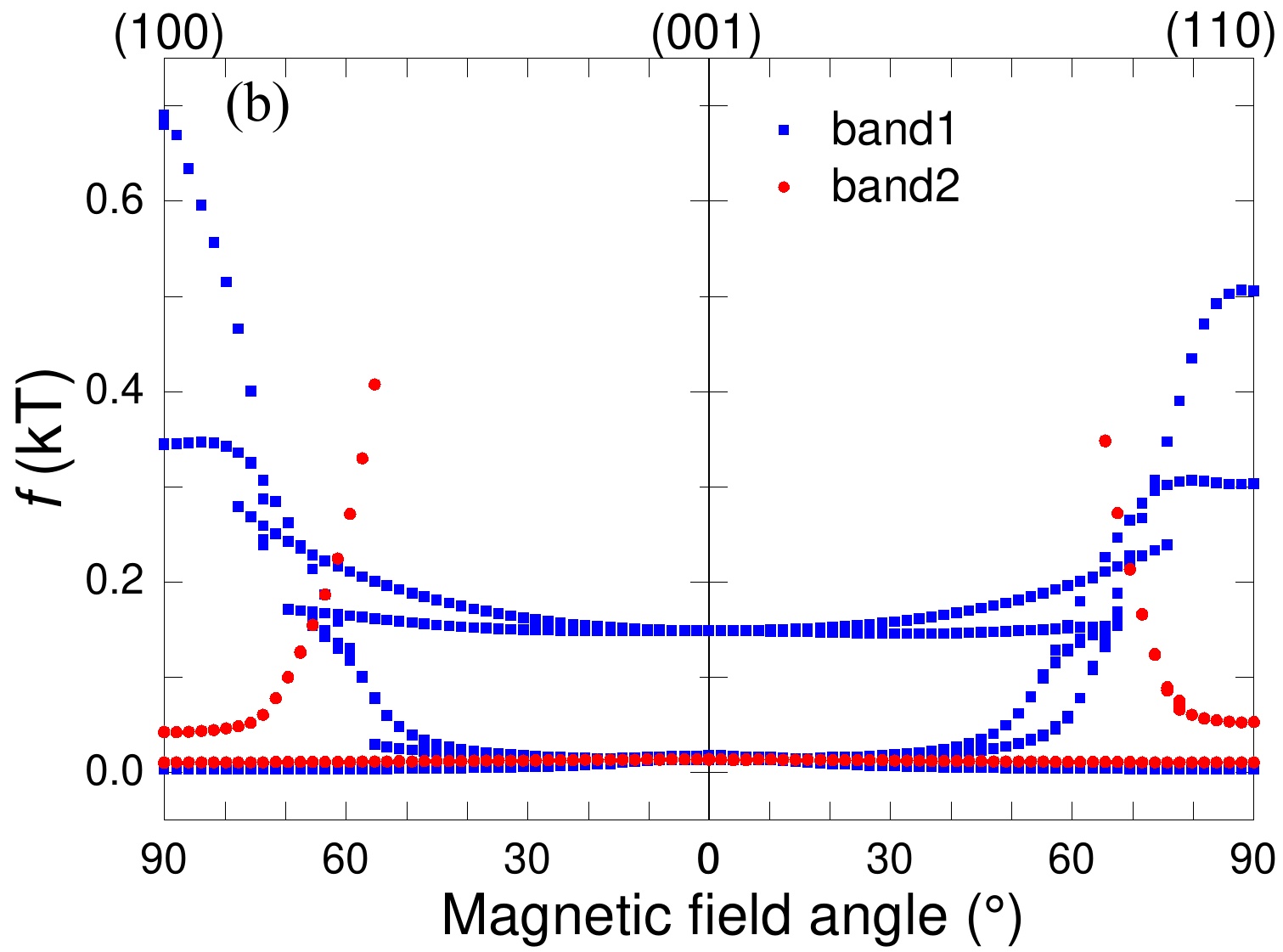} & \includegraphics[width=.49\linewidth,clip]{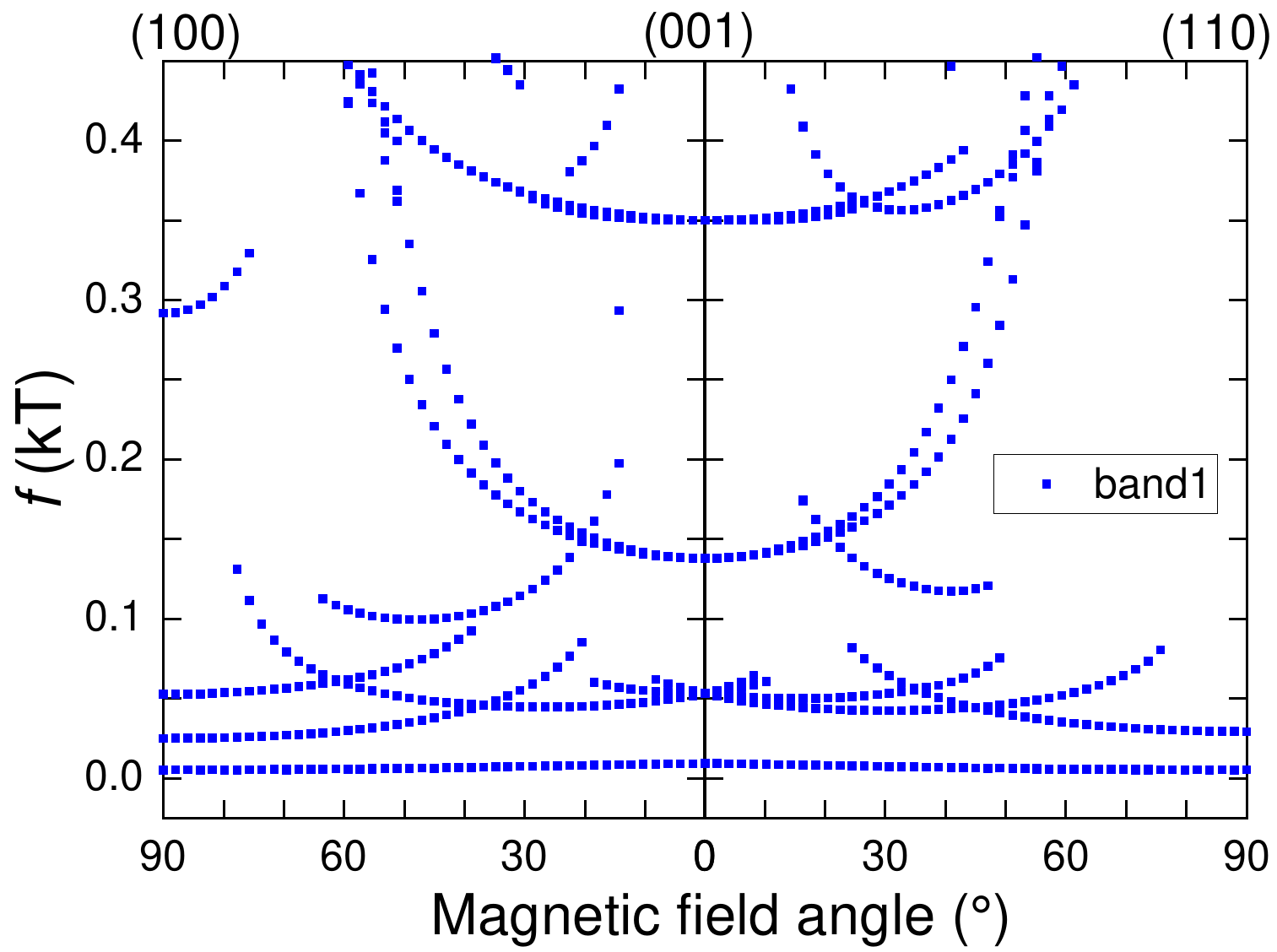}
  \end{tabular}%
  \caption{The calculated dHvA frequencies.
    The top left panel displays the full frequency range, while the bottom left panel focuses specifically on the low-frequency range. The external magnetic field is applied along three different directions: [001] along the hexagonal axis, [100] parallel to the $\Gamma$-$M$ direction, and [110] along the $\Gamma$-$K$ direction.
    The color scheme used in the plot is consistent with Figure 7 of the main text. The exact same plots but for the shifted calculations are shown in the right two panels.
  }
  \label{fig:freq2}
\end{figure}

The density of states (DOS) and their primary orbital/atomic character are shown as a function of energy  in Supplementary Figure~\ref{fig:dos} and Figure~\ref{fig:banddos}.
A more detailed comparison is shown in Figure~\ref{fig:freq2} between the frequencies arising from the Fermi surface for unshifted calculations on the left two panels and calculations in which the Fermi energy was shifted down by 40meV on the right two panels.
\begin{figure}[htb]
	
        \centering
	\begin{tabular}{c}
		\includegraphics[width=.99\linewidth,clip]{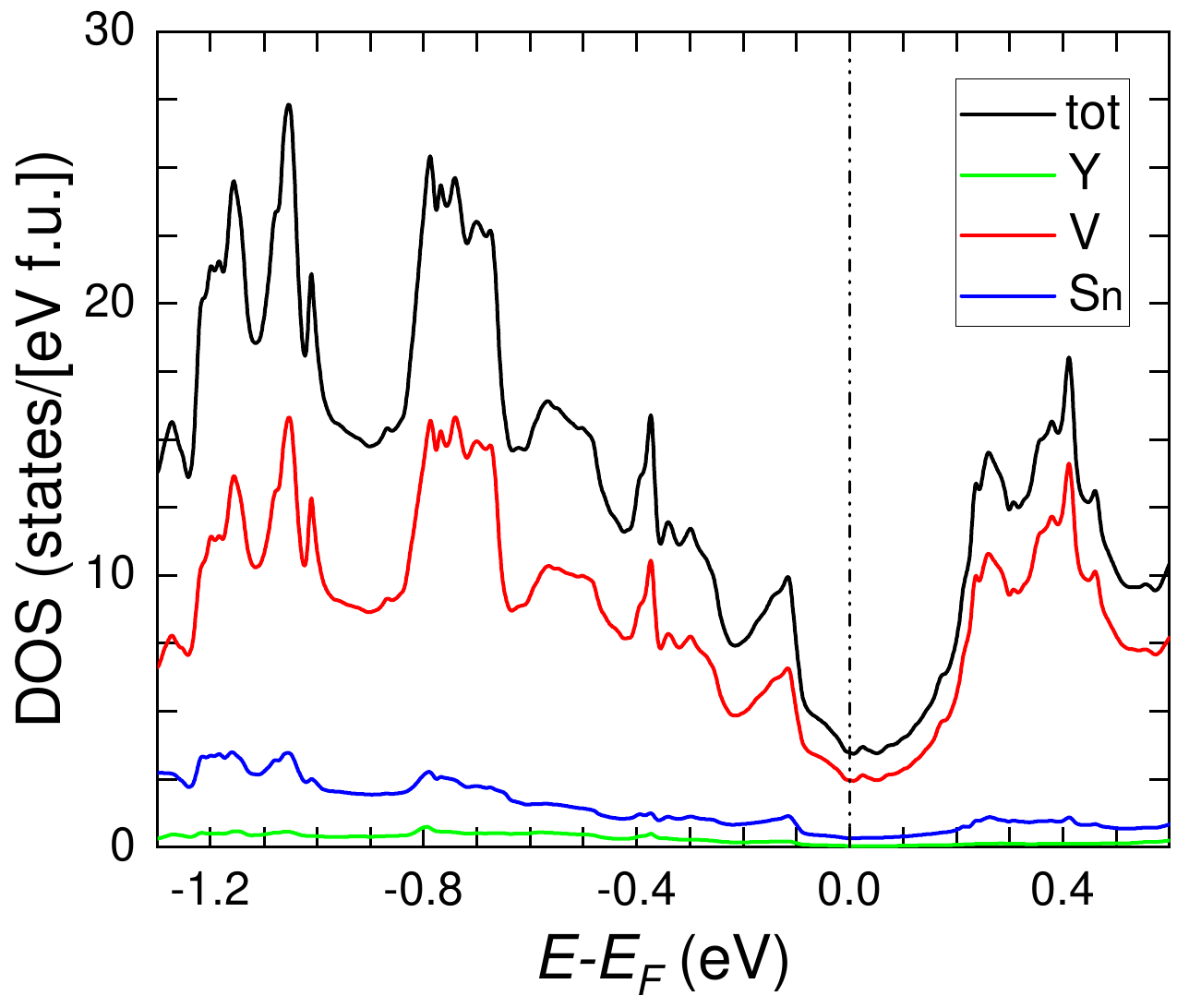}
	\end{tabular}%
	\caption{The total DOS and partial DOS projected on individual elements near the Fermi energy ($E_{ F}$) in \ch{YV6Sn6}.
          The Fermi energy ($E_F$) is at 0 eV. The calculations were performed without SOC. Within the displayed energy window, the contributions of Vanadium (V) sublattices dominate.}
	
	\label{fig:dos}
\end{figure}

\begin{figure}[htb]
	\centering
	\begin{tabular}{c}
		\includegraphics[width=.99\linewidth,clip]{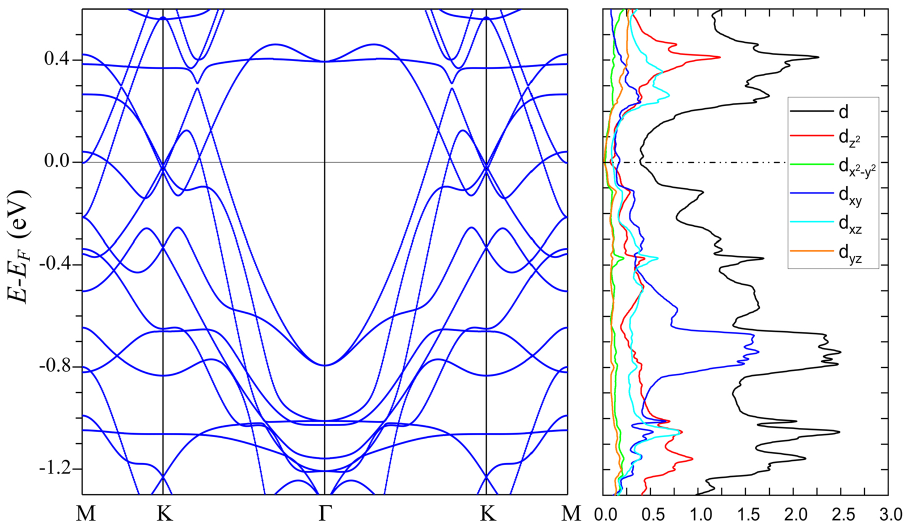}
	\end{tabular}%
	\caption{Scalar-relativistic band structures of \ch{YV6Sn6} in the $k_z$=0 plane and orbital-decomposed $d$-states DOS of Vanadium atoms.
          The peaks in the DOS are associated with flat bands (FB) and display the orbital characters of FB.
The higher FB extends over a large k-space and is associated with the $d_{z^2}$ orbital of the V atom. The lower FB is connected to the $d_{xy}$, $d_{xz}$, and $d_{z^2}$ orbitals.
}
	\label{fig:banddos}
\end{figure}

\subsection{Closeup to experimental frequencies}
Plots of the pockets' frequency dependences with a 2D like $1/cos(\theta)$ fit for the $\gamma$ pocket, as well as a zoomed in look at the small $\alpha$ pocket with a $1/sin(\theta)$ fit, are shown in Supplementary Figure~\ref{fig:angle}. The offset in the left panel focuses on the $\beta_i$ pockets.
\begin{figure*}
    \centering
    \includegraphics[width=1\textwidth]{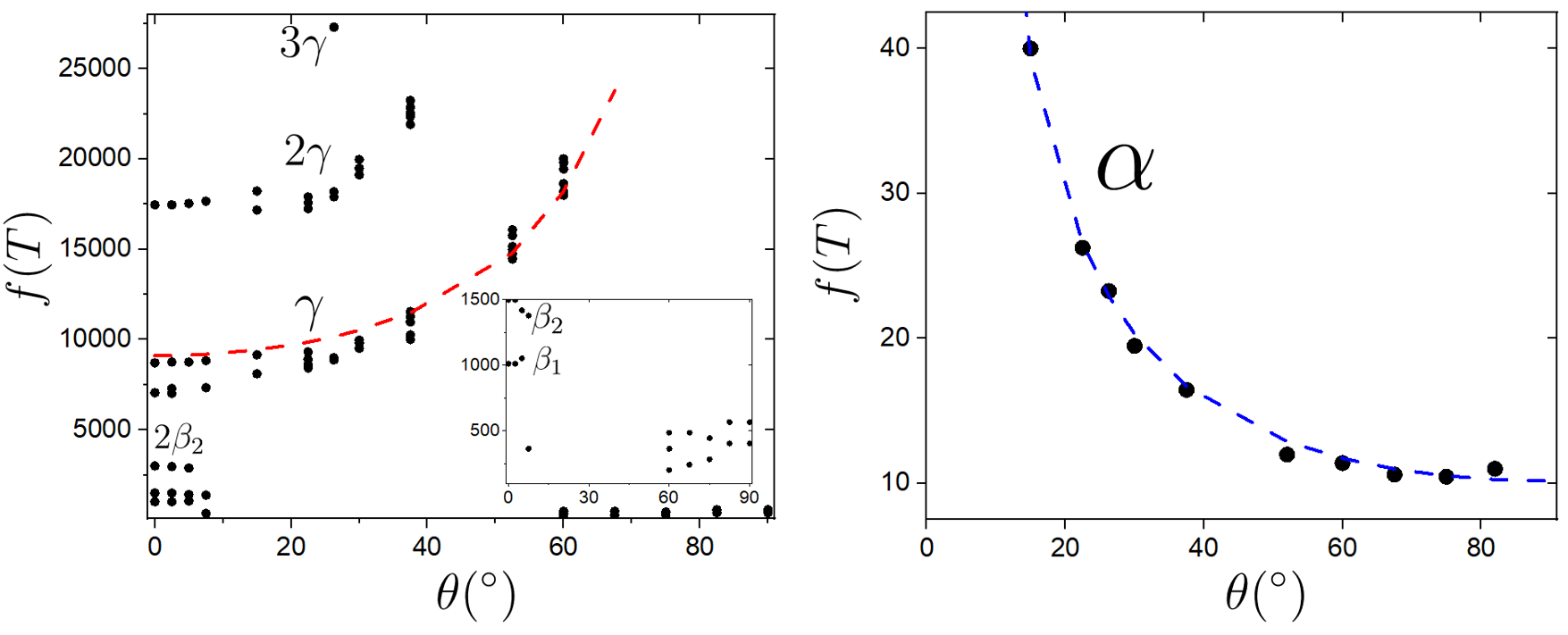}
    \caption{Left panel displays the relevant high frequency peaks of the FFT vs. angle from the c-axis, labeled by the harmonic of the corresponding orbit, with the inset displaying the mid-level frequencies in greater detail. The high frequency peak belonging to the $\gamma$ orbit follows a $1/cos(\theta)$ (red dashed line) relation indicating a 2D like pocket, whereas the low frequency peak belonging to the $\alpha$ orbit (right panel) displays a $1/sin(\theta)$ relation (blue dashed line), indicating a 3D tube like feature.}
    \label{fig:angle}
\end{figure*}
